\documentclass{emulateapj}

\begin{document}

\renewcommand{\topfraction}{1.0}
\renewcommand{\bottomfraction}{1.0}
\renewcommand{\textfraction}{0.0}

\shorttitle{Multiple Star Catalog}
\shortauthors{Tokovinin}

\title{The Updated Multiple Star Catalog}

\author{Andrei Tokovinin}
\affil{Cerro Tololo Inter-American Observatory, Casilla 603, La Serena, Chile}
\email{atokovinin@ctio.noao.edu}

\begin{abstract}
The  catalog  of  hierarchical  stellar  systems with  three  or  more
components is an update of the  original 1997 version of the MSC.  For
two thousand hierarchies, the  new MSC provides distances, component's
masses and periods, as  well as supplementary information (astrometry,
photometry, identifiers,  orbits, notes).  The MSC  content and format
are explained,  its incompleteness and  strong observational selection
are  stressed.  Nevertheless,  the  MSC can  be  used for  statistical
studies  and it  is a  valuable  source for  planning observations  of
multiple stars. Rare  classes of stellar hierarchies found  in the MSC
(with  6  or 7  components,  extremely  eccentric  orbits, planar  and
possibly    resonant   orbits,    hosting    planets)   are    briefly
presented.  High-order hierarchies  have  smaller velocity  dispersion
compared to triples  and are often associated with  moving groups. The
paper  concludes  by  the  analysis   of  the  ratio  of  periods  and
separations between inner and  outer subsystems.  In wide hierarchies,
the ratio  of semimajor axes, estimated  statistically, is distributed
between 3 and 300, with no evidence of dynamically unstable systems.
\end{abstract} 

\maketitle

\section{Introduction}
\label{sec:intro}

This paper contains a collection of observational data on hierarchical
multiple stars and  updates the Multiple Star  Catalog, MSC \citep{MSC}.
The usefulness of  such compilation is supported by   210 citations
to  the  original  paper.  Hierarchical  stellar  systems  are
interesting for several reasons:  as clues to the formation mechanisms
of  multiple  stars,  sites  of interesting  dynamical  phenomena,  or
progenitors  of some  rare  products of  stellar  evolution like  blue
stragglers.   Recent discoveries  of planets  in  hierarchical systems
created  additional  interest  in  such  objects  and  a  stimulus  to
understand the  common origin and  evolution of stellar  and planetary
systems.

The content  of the  MSC results from  random discoveries and  gives a
distorted  reflection of  the real  statistics in  the field.   On the
other hand, volume-limited samples \citep[e.g.][]{R10} are necessarily
small  and contain only  a modest  number of  hierarchies, diminishing
their statistical value. For solar-type stars, an effort to extend the
distance  limit to  67\,pc while  still controlling  the observational
selection has been made by the  author \citep{FG67} and has led to the
first reliable estimate of the frequency of various hierarchies in the
field.  But  even this volume-limited sample  of $\sim$500 hierarchies
is too  small for finding rare  and most interesting  objects, such as
close   tertiary   companions    orbiting   {\it   Kepler}   eclipsing
binaries. Hence the utility of this compilation.

Like its predecessor, the updated MSC can be a source of observational
programs related to multiple stars.  For example, visual binaries with
spectroscopic  subsystems  can   be  selected  for  observations  with
long-baseline  interferometers to determine  orientation of  the inner
orbits  and  the  character  of dynamical  evolution  \citep{Mut2010}.
Tertiary  components  can  be  monitored spectroscopically  to  detect
subsystems or planets.   Although the MSC does not  represent a clean,
volume-limited sample, some statistical inferences can nevertheless be
made using this catalog.

The nature of the data incorporated  in the MSC is briefly outlined in
\S~\ref{sec:data}. The structure and  content of the MSC are explained
in \S~\ref{sec:MSC}.  The  following \S~\ref{sec:zoo} is a  tour of
the multiple-star ``zoo'' highlighting specific members of this
class.  Then in \S~\ref{sec:stat} one statistical aspect of the MSC,
namely the ratio of periods and separations, is presented. The paper
concludes by the short discussion in \S~\ref{sec:disc}.

\section{Data sources and restrictions}
\label{sec:data}

\subsection{Recent literature}
\label{sec:lit}

The MSC was updated for the  last time in 2010, stimulated by the work
of \citet{Eggleton2009} on the  multiplicity of bright stars. However,
more than half  of the hierarchical systems within  67 pc \citep{FG67}
were  discovered  after 2010.   So,  as  a  first step,  those  nearby
hierarchies were added to the  MSC, while its format has been slightly
changed in the process. 

The updated MSC reflects the  results of large multiplicity surveys of
the last decade,  ranging from low-mass and substellar  systems in the
solar  neighborhood   \citep{Dieterich2012,Law2010}  to  A-type  stars
within  75\,pc  \citep{DeRosa2014}  and  massive O-type  stars  beyond
1\,kpc \citep{Sana2014}.  New hierarchies have also been discovered by
speckle interferometry \citep[e.g.][]{SOAR}.

Precise photometry from  space furnished by the {\it  Kepler} and {\it
  COROT} satellites has led to the discovery of thousands of eclipsing
binaries (EBs).  Some  of those EBs show cyclic  variations of eclipse
time  and/or   a  precession  caused  by   relatively  close  tertiary
components  \citep{Borkovits2016}. About 200  such new  compact triple
systems have been added to the MSC.  Ongoing searches of exoplanets by
radial velocity  and transits  also have contributed  new hierarchical
systems, some of those containing planets.

\subsection{Data mining}
\label{sec:mining}

Nowadays  targeted  surveys are  supplemented  by  the  large body  of
all-sky catalogs  available on-line for  data mining.  The  updated MSC
contains  only about  2000  systems,  still a  tiny  number by  modern
standards. Its  content relies  on the four  major catalogs  of binary
stars  presented below,  namely  WDS,  INT4, VB6,  and  SB9.  These
catalogs  are  frequently updated,  so  we  used  their recent  (2016)
on-line versions.

The Washington Double Star Catalog, WDS \citep{WDS} lists thousands of
resolved stellar systems with  two, three, or more components. However,
a substantial fraction  of the WDS entries are  random combinations of
background  stars (e.g.  optical  pairs). Almost  any bright  star has
faint optical components  in the WDS. Optical pairs  are more frequent
in crowded  regions of the  sky; typically their  secondary components
are faint.  Real (physical)  pairs with substantial proper motion (PM)
can    be     distinguished    from    the     optical    ones    (see
\S~\ref{sec:phys}).   This naturally  favors nearby  and low-mass
stars, while  the nature of  distant and massive visual double stars with
small  PMs remains uncertain.  Moreover, several  ``multiple systems''
with common PM  found in the WDS are simply  groups of stars belonging
to  the  same cluster.   Apart  from  the  optical pairs,  the  WDS
contains  spurious pairs that  have not  been confirmed  by subsequent
observations.   For example, many   Tycho  binaries (discovery
codes  TDS and  TDT) with  separations on  the order  of  0\farcs5 are
spurious because they are not confirmed by speckle interferometry. 

Considering this ``noise'', no  attempt to extract triple systems from
the WDS has been made  in the original MSC compilation.  Now candidate
triples  with  PMs  above  50   mas~yr$^{-1}$  were  selected  from  the  WDS
automatically. Some of them passed  the reality test and were added to
the new  MSC.  The Fourth  Catalog of Interferometric  Measurements of
Binary Stars, INT4 \citep{INT4} was consulted in the process.

Hierarchical  systems  with  known  orbits are  of  special  interest.
Entries of the Sixth Catalog  of Visual Binary Orbits, VB6 \citep{VB6}
were  cross-checked  with  the  WDS  for the  presence  of  additional
physical  components.  Similarly, the  Ninth Catalog  of Spectroscopic
Binary Orbits,  SB9 \citep{SB9} was  matched with the WDS.   This data
mining has improved the census of triple systems among ``orbital''
binaries.   The number of those binaries  is steadily growing,
while not all of them are featured in SB9 and VB6. Therefore, scanning
of  the current  literature is  an  essential complement  to the  data
mining.

\subsection{Limitations}
\label{sec:lim}

The  vast majority  of hierarchical  systems  in the  MSC are  located
within 1\,kpc from the Sun.  Nearby objects are generally brighter and
have the advantage  of being resolvable; their large PMs help  to get rid of
optical  companions.  Although  modern observations  go very  deep and
far,  we took  the decision  to ignore  hierarchies in  the Magellanic
clouds and  beyond, as  well as many  distant massive stars. 

Some eclipsing  binaries have  periodic eclipse time  variations (ETV)
that can be explained either by presence of a tertiary component or by
magnetic  cycles in  one  or  both components  of  the eclipsing  pair
\citep[see  the  discussion  in   ][]{Liao2010}.  In  many  cases  the
existence  of  tertiary  components  is  confirmed  by  other  methods
(e.g.  J02091+4088  or BX~And,  where  the  62  year tertiary  has  an
astrometric  orbit).  On  the  other hand,  some  tertiary  components
discovered by ETV  remain controversial.  For example, \citet{V776Cas}
claim that J01534+7003 (V776~Cas) is  a quadruple system, based of the
ETV with a  period of 23.7 years, of which one  cycle is observed. The
implied  tertiary  component  of  one  solar mass  with  a  fractional
luminosity  of  0.15  should  be  detectable in  the  spectrum.   Yet,
\citet{D'Angelo2006}  found  only a  weak  spectral  signature with  a
fractional luminosity of 0.015 corresponding to the visual companion B
at 5\farcs4 separation.  Accordingly, V776~Cas is listed in the MSC as
triple,  rather than  quadruple, until  the reality  of the  23.7 year
subsystem  is confirmed.   It is  noteworthy that  tertiary components
found by ETV usually have circular orbits, while typical binary orbits
have  non-zero  eccentricity.   We  leave aside  many  triple  systems
discovered by ETV until they are confirmed by other techniques.

For most  systems, the MSC does not  provide bibliographic references.
Given the coordinates and/or  common identifiers, the bibliography can
be  retrieved  from  Simbad.   Information  on  resolved  binaries  is
retrieved from  the WDS  and INT4, the  orbital elements from  VB6 and
SB9.  The notes give  references where appropriate, e.g for subsystems
that are not  found in the main catalogs.   Additional parameters such
as masses and periods are estimated by the methods explained below.

\begin{figure}
\epsscale{1.0}
\plotone{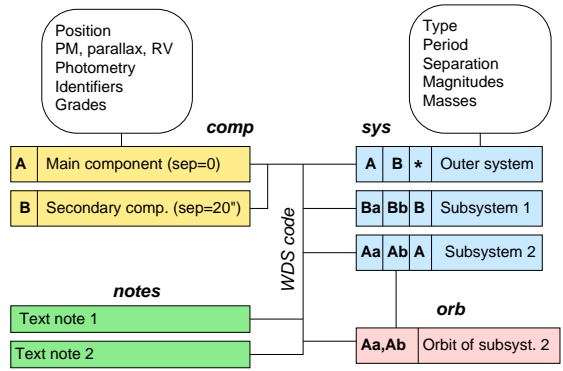}
\caption{Structure of the  MSC. It consists of four  tables {\tt comp,
    sys, orb, notes} linked by  the WDS code, unique for each multiple
  system.
\label{fig:db}  }
\end{figure}

\section{The new MSC}
\label{sec:MSC}

\subsection{Catalog structure}
\label{sec:cat}

The  structure of the  MSC has  been described  in the  original paper
\citep{MSC};  it changed  only  slightly.  The  MSC  consists of  four
tables linked by  the common field based on  the WDS-style coordinates
for the  J2000 epoch (Figure~\ref{fig:db}).   We call them  WDS codes,
although,  strictly  speaking, the  equivalence  applies  only to  the
resolved visual  binaries actually listed  in the WDS. The  tables are
available in full  electronically as text files. They  are too wide to
fit in the printed page, therefore providing their fragments would not
be  useful to the  reader. Instead,  we describe  the content  of each
table.    The format  codes indicate  types of  the fields  in the
  machine-readable tables  (A --  string, F --  floating number,  I --
  integer number).

Part  of the MSC  content is  available in  the on-line  database that
allows   searches  on   identifiers  or   coordinates.\footnote{  {\url
    http://www.ctio.noao.edu/\~{}atokovin/stars/} }  The same
  link also points to the full ASCII tables and  old versions of the MSC.

\begin{figure}
\epsscale{1.1}
\plotone{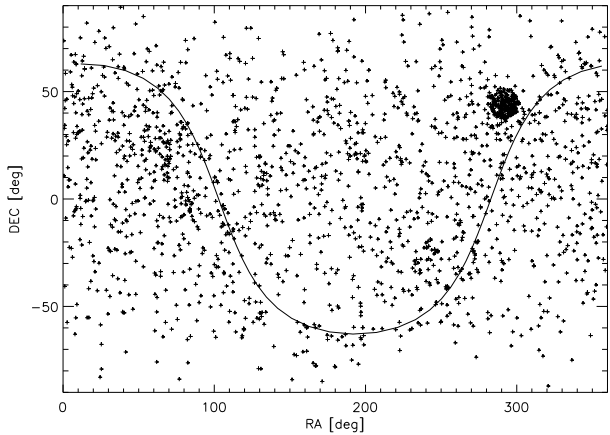}
\caption{Distribution of  the MSC objects  in the sky. The  line shows
 the Galactic equator.
\label{fig:skymap}  }
\end{figure}

Illustrating  the  MSC   content,  Figure~\ref{fig:skymap}  plots  its
distribution  on the sky  in equatorial  coordinates.  The  cluster of
points at  $(290\degr, +50\degr)$  corresponds to multiple  systems in
the {\it  Kepler} field. Even  without those multiples, the  number of
systems  with  northern declinations  is  larger  than  the number  of
southern  multiples, reflecting the  historical bias  in favor  of the
northern sky.

\begin{figure}
\epsscale{1.0}
\plotone{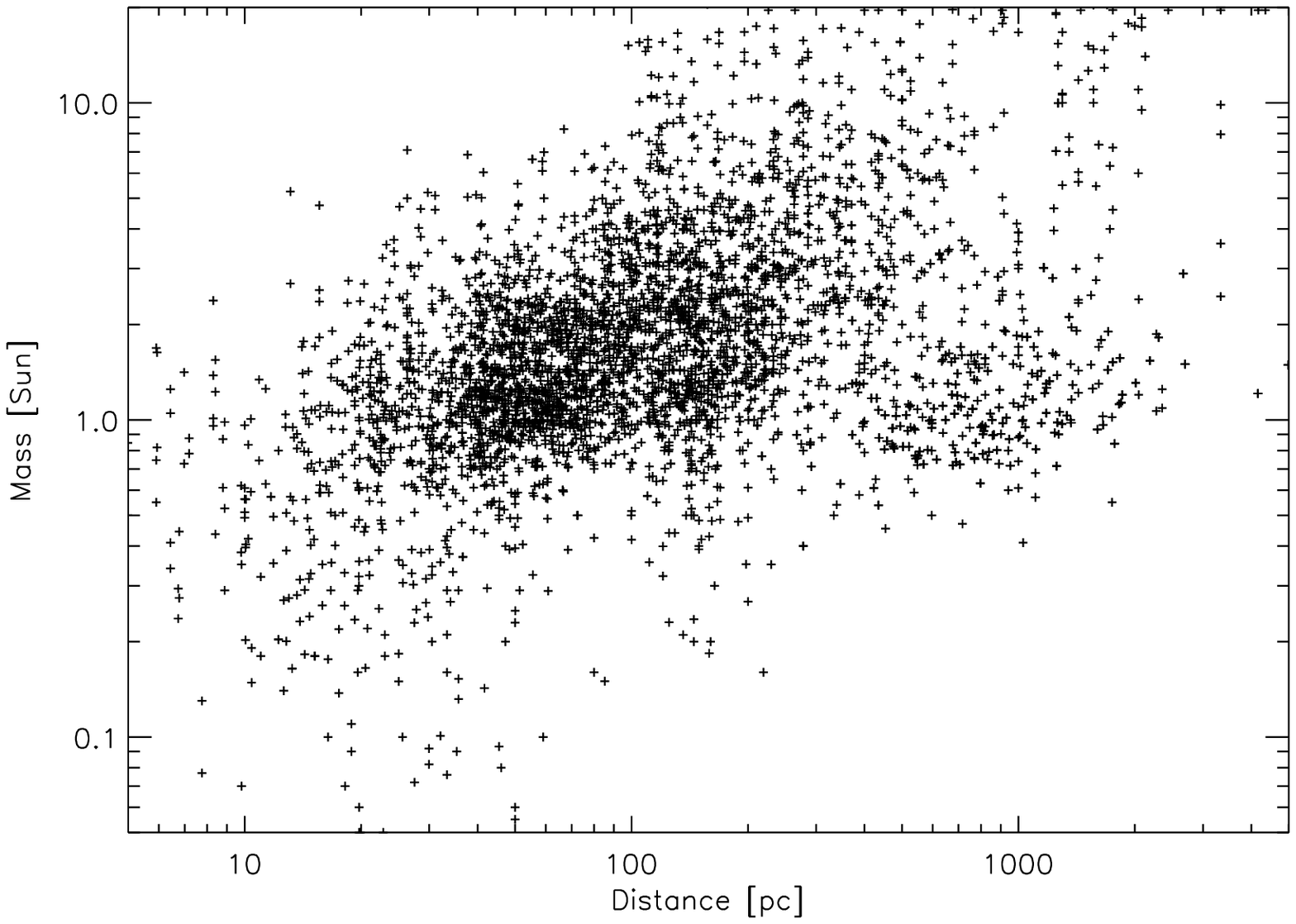}
\plotone{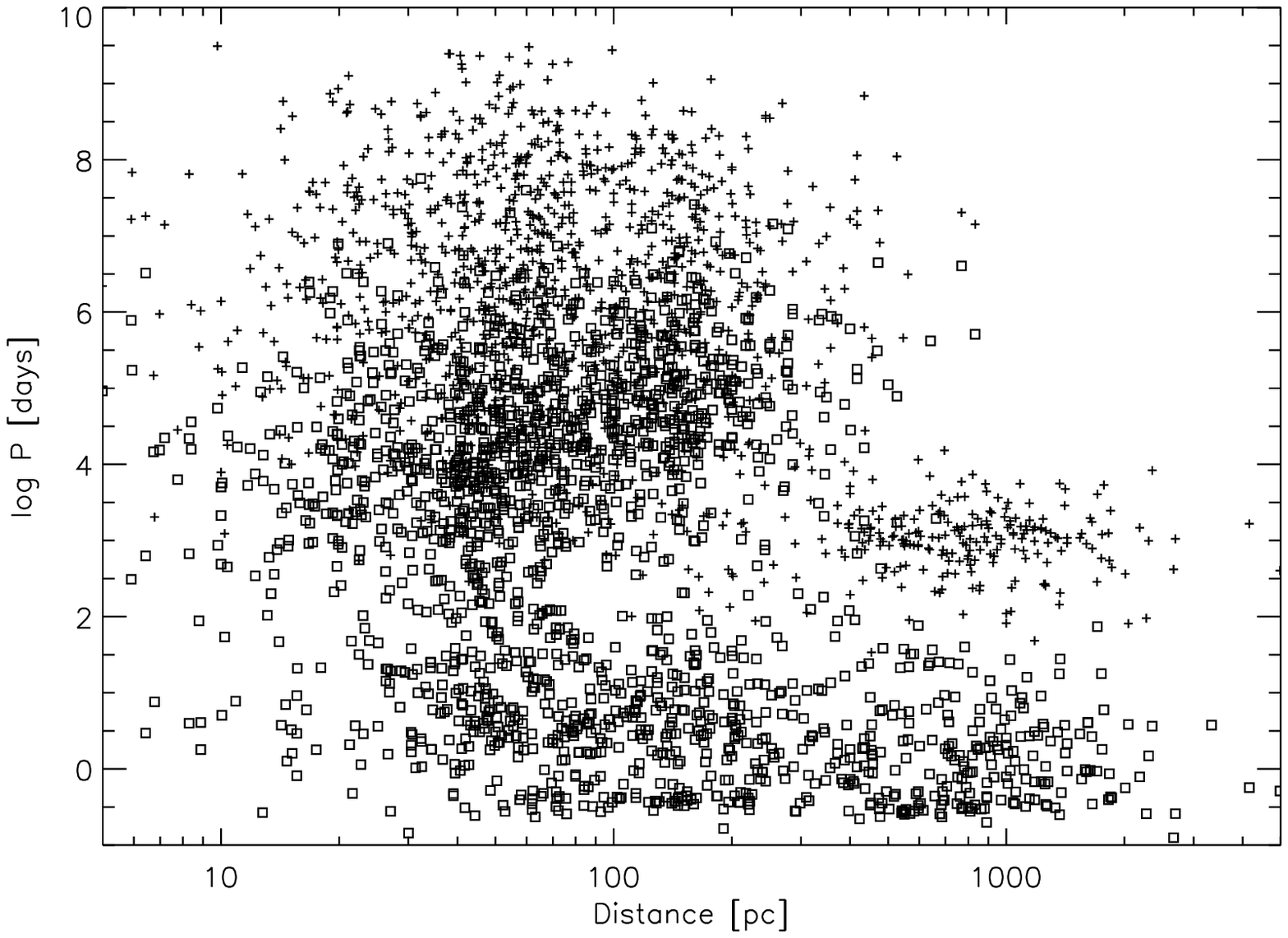}
\caption{MSC content vs. distance. Top: primary mass vs. distance.
Bottom: periods vs. distance for primary mass less than 3 ${\cal
  M}_\odot$ (crosses -- outer periods, squares -- inner periods). 
\label{fig:dist}  }
\end{figure}

Figure~\ref{fig:dist} illustrates dependence of the MSC content on the
distance $d$;  the median distance  is 113\,pc.  The upper  plot shows
strong correlation of  the primary mass with the  distance; only a few
low-mass  multiples are discovered  beyond 30\,pc.   Not surprisingly,
the number of objects in the  MSC is not proportional to $d^3$, as one
might expect  for a spatially  uniform population. Instead,  the $d^2$
law is a good  fit to the cumulative counts at $d  < 50$\,pc. When the
range of masses is restricted to solar-type stars, the $d^2$ law still
holds   well,  demonstrating   incompleteness of the MSC  even   for  those
well-studied objects.   The lower plot of periods  vs.  distance gives
an evidence  of a ``gap'' at  $P \sim 100$ days,  possibly a selection
effect.  The lack of long periods (wide binaries) beyond $\sim$300\,pc
is also obvious;  such binaries have small PMs  and are not discovered
by  the current  surveys.  The  {\it Kepler}  eclipsing  binaries with
tertiary components and outer periods around 1000 days make a distinct
group   at    the   distance   of   $\sim$1\,kpc.     The   plots   in
Figure~\ref{fig:dist} give  an idea  of the ``typical  `` hierarchical
system: it has the primary mass between 0.5 and 3 ${\cal M}_\odot$ and
is located within 300\,pc from the Sun.

\begin{deluxetable}{l  l l }
\tabletypesize{\scriptsize}
\tablewidth{0pt}
\tablecaption{Components table (comp) \label{tab:comp}}
\tablehead{
\colhead{Field} &
\colhead{Format} &
\colhead{Description} 
}
\startdata
WDS    & A10 & WDS code (J2000)  \\
RA     & F10.5 & Right ascension J2000  (degrees) \\
DEC    & F10.5 &  Declination J2000  (degrees) \\
Parallax & F8.2 &  Parallax  (mas) \\
Refplx   & A4   &   Reference code for parallax\tablenotemark{a} \\
PMRA     & F7.1 & Proper motion in RA (mas~yr$^{-1}$) \\
PMDE     &  F7.1 &  Proper motion in Dec (mas~yr$^{-1}$) \\
RV       & F6.1  &  Radial velocity (km~s$^{-1}$) \\
Comp     & A2    &  Component label \\
Sep      &  F7.1 &  Separation  (arcsec) \\
Sp      & A8     &  Spectral type \\
HIP     & I6     &  Hipparcos number \\
HD      &  I6    &  HD number \\
Bmag    & F5.2   &  $B$-band magnitude \\
Vmag    & F5.2   &  $V$-band magnitude \\
Imag    & F5.2   &  $I_{\rm C}$ band magnitude \\
Jmag    & F5.2   &  $J$-band magnitude \\
Hmag    & F5.2   &  $H$-band magnitude \\
Kmag    & F5.2   &  $K$-band magnitude \\
Ncomp  & I2      & Number of physical components \\
Grade  & I1      &  Grade\tablenotemark{b} \\
Ident  &  A40    & Other identifiers      
\enddata
\tablenotetext{a}{Parallax codes: 
HIP -- Hipparcos, 
Gaia -- Gaia DR1, 
dyn -- dynamical,
orb -- orbital,
pN -- photometric from component N, 
bib -- taken from literature.}
\tablenotetext{b}{See \S~\ref{sec:grades}.}
\end{deluxetable}

\begin{deluxetable}{l  l l }
\tabletypesize{\scriptsize}
\tablewidth{0pt}
\tablecaption{Systems table (sys)  \label{tab:sys}}
\tablehead{
\colhead{Field} &
\colhead{Format} &
\colhead{Description} 
}
\startdata
WDS    & A10 & WDS code (J2000)  \\
Primary   &  A3 &  Primary component label \\
Secondary &  A3 &  Secondary component label \\
Parent &  A3    & Parent label\tablenotemark{a} \\
Type   &  A6    &  Observing technique/status\tablenotemark{b} \\
P      &  F10.4 & Orbital period \\
Punit &   A1    & Units of period\tablenotemark{c} \\
Sep   &  F8.3   & Separation or semimajor axis \\
Sepunit &  A1   & Units of separation\tablenotemark{d} \\
Pos. angle &  F5.1 & Position angle (deg) \\
Vmag1 &   F5.2   & $V$-band magnitude of the primary \\
Sp1   &  A5      & Spectral type of the primary \\
Vmag2 &   F5.2   & $V$-band magnitude of the secondary \\
Sp2   &  A5      & Spectral type of the secondary \\
Mass1 &  F6.2    & Mass of the primary (sun) \\
Mcode1 &  A1     & Mass estimation code for Mass1\tablenotemark{e}  \\
Mass2 &  F6.2    & Mass of the secondary (sun) \\
Mcode2 &  A1     & Mass estimation code for Mass2 \tablenotemark{e}  \\
Comment &  A20   & Comment on the system 
\enddata
\tablenotetext{a}{Parent points to the component identifier in a
     higher-level system to which the current system belongs and thus
     codes the hierarchy by reference. Two special symbols are used:
       * means root (system at the highest hierarchical level);
       t means trapezium-type, non-hierarchical system.
}
\tablenotetext{b}{See Table~\ref{tab:type}.}
\tablenotetext{c}{Period units: d -- days, y -- years, k -- kiloyears,
  M -- Myrs.}
\tablenotetext{d}{Separation units: `` -- arcsec, ' -- arcmin, m -- mas.}
\tablenotetext{e}{ Mass codes:
          r -- given in the original publication,
          v -- estimated from absolute magnitude,
          a -- estimated from spectral type or color index,
          s -- sum of masses for the sub-system(s),
          q -- estimated from primary mass and mass ratio of SB2,
          m -- minimum secondary mass for SB1.
}
\end{deluxetable}

\subsection{Components table (comp)}
\label{sec:comp}

The main table {\tt comp}  contains data on the individual components,
both  primary and secondary:  astrometry, photometry,  and identifiers
(see  Table~\ref{tab:comp}).  The  MSC  does not  provide errors  of
astrometry  or photometry, as  these data  are recovered  from various
heterogeneous  sources; it  should not  be used  as an  astrometric or
photometric catalog.

The  brightest star in  each multiple  system --- its  primary
component ---
always has  an entry  in the {\tt  comp} table. Other  components with
separations larger than a few arcseconds from the primary, if present, have
their own entries.  The  non-zero separation distinguishes them from
the  primaries.  To  count multiple  systems, only  primary components
with  zero  separation  should  be considered.   However,  photometry,
astrometry   and  identifiers  of   the  secondary   components,  when
available,  are  very useful  for  evaluating  their  relation to  the
primary and for compilation of observing programs. 

The unknown (missing) parameters in the MSC  have zero values. This
feature is inherited from the original MSC and should be kept in mind
when using the tables.  

The MSC provides the HD and  HIP numbers for locating the stars in the
SIMBAD. However, the  objects in the new MSC  are, on average, fainter
than in  the old one, and  an increasing fraction  of them (especially
the secondary components) lack  traditional identifiers.  On the other
hand, faint  stars may  have a variety  of useful aliases,  e.g.  {\it
  Kepler}  or  2MASS  numbers,  variable-star designations,  etc.   In
response  to this  situation, the  new  MSC contains  a collection  of
arbitrary identifiers in free  format.  Coordinates and identifiers
help to  retrieve information on  components from {\it  Vizier} or
other sources.

\subsection{Systems table (sys)}
\label{sec:sys}

\begin{figure}
\epsscale{1.0}
\plotone{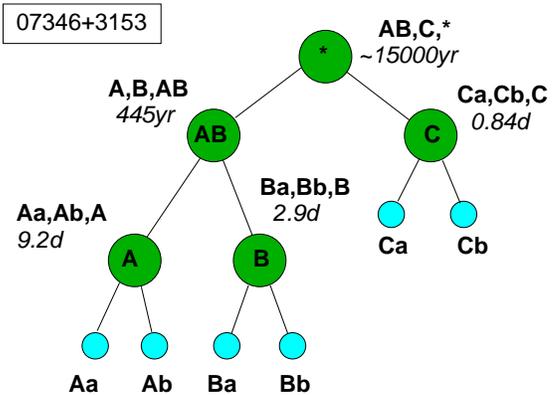}
\caption{Hierarchical   structure  of   the  sextuple   system  Castor
  (J07346+3153,  $\alpha$ Gem). Large  green circles  depict subsystems
  (their  designations  and periods  are  given),  small blue  circles
  denote individual stars.
\label{fig:Castor}  }
\end{figure}

\begin{deluxetable}{l  l }[ht]
\tabletypesize{\scriptsize}
\tablewidth{0pt}
\tablecaption{System types \label{tab:type}}
\tablehead{
\colhead{Type} &
\colhead{Discovery technique} 
}
\startdata
C[mhrp] & CPM pair and criteria of relation \\
c       & Wide pair with uncertain status \\
v, o   & Resolved visual or occultation binary \\ 
V       & Visual binary with known orbit \\
a       & Acceleration binary \\
A       & Astrometric binary with known orbit \\
s, s2   & Spectroscopic binary (e.g. double-lined) \\
S1, S2  & Spectroscopic binary with known orbit \\
e, E     & Eclipsing binary \\
E*      & Eclipse time variations \\
X[mhrp]  & Spurious pair (e.g. optical)        
\enddata
\end{deluxetable}

The table {\tt sys} contains information on the individual subsystems:
their    types,     periods,    separations,    and     masses    (see
Table~\ref{tab:sys}).  The fields  {\it primary}, {\it secondary}, and
{\it parent} contain component's identifies and, jointly, describe the
hierarchical      structure.      This      is      illustrated     in
Figure~\ref{fig:Castor}  using  the   sextuple  system  Castor  as  an
example. The outer hierarchical level (root) is the wide binary AB,C,*
consisting of the  components AB and C (root is  coded by the asterisk
in  the parent  field).   The  component C  is  a close  spectroscopic
binary.   The  system  A,B  is  a  visual  binary  with  a  period  of
445\,years.   In  turn,  both  its   components  A  and  B  are  close
spectroscopic binaries.  So, a  component of the hierarchy may contain
several  stars.   To avoid  confusion,  in  most  cases the  component
designations in  the MSC  match those in  the WDS,  although arbitrary
character strings can  be used as component designations  just as well
and there are no strict  rules here. For example, a resolved secondary
can be  designated as Ba,Bb,B or as B,C,BC  both in the MSC  and in the
WDS.  Systems are designated  by their primary and secondary component
(and,  sometimes,  root)  joined  with  a  comma.   This  notation  is
explained in \citep{Tok2005,Tok2006}.

Although the MSC  contains predominantly hierarchical stellar systems,
in  several  cases the  separations  between  resolved components  are
comparable and  it is not known  if those systems  are hierarchical or
not.  Such apparently  non-hierarchical configurations are called {\it
  trapezia}.  In  the MSC, trapezia are  denoted by the  symbol 't' in
the parent field instead of the  usual '*'.  Thus, a trapezium has two
or more upper-level  systems with the parent 't',  but no root system.
Note  that the  Orion  Trapezium which  gave  name to  this class  of
objects is in fact a young stellar cluster.  We do not consider it as
a single entry,  but include three hierarchies belonging  to the Orion
Trapezium as separate entries, each with its own WDS code.

The types  of the  systems reflect the  discovery techniques,  with an
obvious   coding:    C   --    common   proper   motion    (CPM,   see
\S~\ref{sec:phys}),  v -- visual,  etc. (see  Table~\ref{tab:type}).  A
system can  have several types, e.g.  a  resolved spectroscopic binary
has types v and s (or V  and S if both visual and spectroscopic orbits
are known).  The special  type X denotes  optical or  spurious systems
that are kept in the {\tt sys} table only for completeness.

The type defines the meaning of the period and separation. For systems
with known orbits (types A, V,  S, E), those fields contain the actual
period  and the  semimajor axis  $a$ in  angular units.   For resolved
binaries  without known  orbits, the  separation $\rho$  is  listed in
place of the semimajor axis,  while the period $P^*$ is estimated from
the third Kepler law by  assuming that the projected separation equals
the semimajor axis:
\begin{equation}
P^* = (\rho /p)^{3/2} M^{-1/2} ,
\label{eq:Pstar}
\end{equation}
where $M$ is  the mass sum in solar units, $p$  is the parallax, and
$P^*$ is  in years.  The ratio $a  /\rho$ is a random  variable with a
median value close  to one and a typical variation by  a factor of two
around  the  median \citep{FG67}.   Therefore,  the estimated  periods
$P^*$ are typically within a  factor of 3 from the true periods.
The same formula  (\ref{eq:Pstar}) is used to compute  the semimajor axis
of close (unresolved) binaries with known periods.  Unknown periods or
separations have the default zero value.

\subsection{Orbits  and notes}
\label{sec:orb}

\begin{deluxetable}{l  l l }
\tabletypesize{\scriptsize}
\tablewidth{0pt}
\tablecaption{Orbit table (orb). \label{tab:orb}}
\tablehead{
\colhead{Field} &
\colhead{Format} &
\colhead{Description} 
}
\startdata
WDS    & A10 & WDS code (J2000)  \\
System & A8  &  Primary,Secondary labels \\
$P$    &  F12.4 & Orbital period (see Punit) \\
$T_0$  & F10.4  & Periastron epoch \tablenotemark{a} \\
$e$    &  F6.3  & Eccentricity \\
$a$    & F8.4  &  Semi-major axis (arcsec) \\
$\Omega$ & F6.2 & P.A. of the ascending node (deg) \\
$\omega$ & F6.2 & Argument  of periastron (deg) \\
$i$     & F6.2  &  Inclination (deg) \\
$K_1$  &  F6.2  & Semi-amplitude of the primary (km~s$^{-1}$) \\ 
$K_2$  &  F6.2  & Semi-amplitude of the secondary (km~s$^{-1}$) \\ 
$V_0$  &  F8.2  & Center-of-mass velocity (km~s$^{-1}$) \\ 
Node   & A1    & Component to which the node refers \\
Punit  & A1    & Unit of period (days ot years) \\
Comment & A30  & Note (may include bibcode)   
\enddata
\tablenotetext{a}{Besselian year if $P$ in years, JD$-$2400000 if $P$ in days.}
\end{deluxetable}

\begin{deluxetable}{l  l l }
\tabletypesize{\scriptsize}
\tablewidth{0pt}
\tablecaption{Notes table (notes) \label{tab:notes}}
\tablehead{
\colhead{Field} &
\colhead{Format} &
\colhead{Description} 
}
\startdata
WDS    & A10 & WDS code (J2000)  \\
Text & A80   & Text of the note \\
Bibcode & A19 & Bibcode 
\enddata
\end{deluxetable}

The third table {\tt  orb} (see Table~\ref{tab:orb}) lists elements of
visual, spectroscopic,  or combined orbits, when  available.  They are
copied mostly  from the catalogs  of visual and  spectroscopic orbits,
VB6 \citep{VB6} and SB9  \citep{SB9} respectively, with some additions
from the recent literature. For unpublished spectroscopic orbits by
D.~Latham (2012, private communication), only periods are given in the
{\tt sys} table, with the 'Cfa' reference in the comment field. 

Finally, the table  {\tt notes} contains notes (Table~\ref{tab:notes})
in  the free-text  format. The  new field  {\tt bibcode}  is  added to
provide the source  of some notes.  However, in  most cases it remains
empty (we made  no effort to provide bibcodes  for references that are
given  in  the   old  MSC  in  free  format).   The  notes  amply  use
abbreviations (e.g. plx for parallax,  PM for proper motion, etc.) and
short codes for common references, such as R10 for \citep{R10}. A list
of such references is given in \citep[][paper I, Table 1]{FG67}.

\subsection{Masses and distances}
\label{sec:mass}

When the first version of the  MSC was compiled, the distances to most
objects  were  not  measured   directly,  but  rather  estimated  from
photometry and/or spectral types.   The knowledge of distance and mass
is needed  to evaluate  the period from  the projected  separation, or
vice versa. Now the situation  has changed radically, as the distances
to most  hierarchies are measured by {\it  Hipparcos} \citep{HIP2} and
{\it Gaia} DR1 \citep{Gaia}; the new MSC contains about 900 parallaxes
from  each  of those  sources.   However,  {\it  Gaia} does  not  give
parallaxes for stars that are either too bright or non-single, not yet
processed in the DR1.   The DR1  catalog  also provides  relative
positions of some binaries useful for evaluating their motion.

Masses  of the main-sequence  components can  be estimated  from their
absolute  magnitudes more reliably  than from  the colors  or spectral
types. This  is now  the preferred method.   We use the  tabulation of
\citet{Mamajek}, valid  for spectral  types later than  O9V. Uncertain
masses evaluated  from the spectral  types are retained only  for some
stars.  For objects without  trigonometric or dynamical parallaxes, we
evaluate the absolute magnitude either  from the spectral type or from
the  $V-K$  color  (assuming  a  single  main  sequence  star  without
extinction)  and  hence derive  the  photometric distance.   Dynamical
parallaxes estimated from the visual  orbits can be more accurate than
photometric or even trigonometric  parallaxes.  When all methods fail,
a  rough guess  of  the distance  is  made.  Estimates  of masses  and
distances given in the literature, whenever available, are preferred
to our own estimates.
 
If the {\tt comp} table has several entries for the same system, the
distances to the secondary component are assumed to be the same as for
the primary, unless independent measurements for those components are
available.

\begin{figure}
\epsscale{1.0}
\plotone{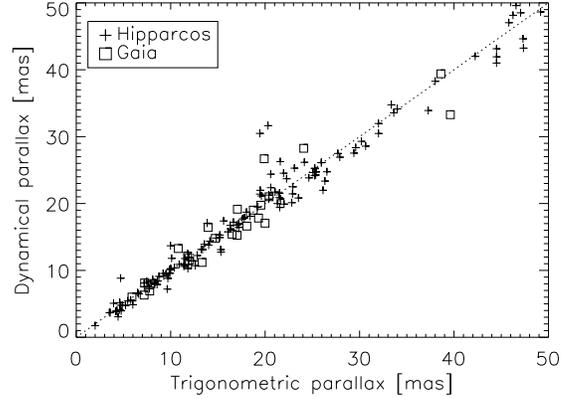}
\caption{Comparison of trigonometric and dynamical parallaxes. The
  dotted line shows equality of parallaxes.
\label{fig:plx}  }
\end{figure}

As  a  consistency   test,  Figure~\ref{fig:plx}  compares  the  known
trigonometric parallaxes  with the dynamical  parallaxes computed from
the visual orbits and the estimated masses.  Only orbits of grade 4 or
better are used,  and the comparison is restricted  to parallaxes less
than 50\,mas.   The 144 binaries with {\it  Hipparcos} parallaxes have
the mean  difference $p_{\rm HIP} -  p_{\rm dyn}$ of  $-0.001$ mas and
the  rms scatter of  1.98\,mas. For  the 23  binaries with  {\it Gaia}
parallaxes,  the mean difference  is $-0.06$\,mas  and the  scatter is
2.57\,mas.   The near-zero  average  difference proves  that the  mass
estimates in  the MSC are  unbiased.  Inspection of  the discrepancies
suggests the reason to be questionable visual orbits (even some orbits
of grade 3), rather than the errors of masses or parallaxes.

\subsection{Physical or optical?}
\label{sec:phys}

Most visual binaries  wider than 1\arcsec ~have the  type 'C', meaning
common  proper motion (CPM).   Several classical  methods are  used to
distinguish  real   (physical)  components  from   chance  projections
(optical).  In the {\tt sys}  table, each method has its corresponding
flag after  the letter C.  The similarity  of PMs, flag 'm', is a
useful   indicator    when   the    PMs   are   large    enough   (say
$>$30\,mas~yr$^{-1}$) to  distinguish both stars  from the background.
Several  systems  with  small  PMs,  listed in  the  original  MSC  as
physical, turned  out to be  chance projections.  A  related criterion
uses  the relative  angular motion  between the  components  $\mu$ (in
mas~yr$^{-1}$).  Bound  binaries cannot move too  fast. This condition
can be expressed as a limit on the parallax $p$,
\begin{equation} 
p > p_{\rm crit} = (\rho \mu^2)^{1/3} (2 \pi)^{-1/3} M^{1/3},  
\label{eq:plx}
\end{equation}
where $M$  is the  mass sum in  solar units  \citep[see e.g.][]{TK16}.
This equation  can be re-cast as the upper limit  $\mu_{\rm crit}(p, M)$.
Simulations  show  that the  typical  projected  relative velocity  is
$\sim$1/3   of   its   critical   value  $\mu_{\rm   crit}$   given   by
(\ref{eq:plx}).   This  leads  to  the  statistical  estimate  of  the
parallax  $p \approx  0.26  (  \rho \mu^2  M)^{1/3}$,  similar to  the
hypothetical parallax  in the sense  of \citet{Ressell}, $p_{\rm  h} =
0.418  (\rho   \mu^2)^{1/3}$.   Considering  the   randomness  of  the
instantaneous orbital motions and the inevitable measurement errors of
$\mu$, the  hypothetical parallax must not exceed  the actual parallax
by more than three times, as a rule of thumb.  This criterion, denoted
by the  flag 'h', depends on  the accuracy of  relative positions
used to  compute $\mu$, as well  as on the  time base.  Unfortunately,
the measurements listed in the  WDS (especially the first ones) can be
inaccurate, compromising  this criterion. We  evaluate subjectively if
the observed  displacement of the binary  reported in the  WDS is real
and, if so, use the hypothetical parallax to reject optical pairs that
move  too fast.   Uncertain pairs  have a  question mark  in  the type
field.

Relative positions of  wide pairs are not measured  or cataloged with
sufficient accuracy for evaluating  their slow relative motion, making
the hypothetical parallax  useless. However, when the PM  is large and
the relative displacement between  the first and the last measurements
is much  less than implied  by the PM,  the stability of  the relative
position indicates  that the  pair is physical.   Such cases  are also
denoted by the flag 'h'. 

Common  RV  (flag 'r')  is  another  independent criterion  of  a
physical binary. It can be falsified by large measurement errors or by
the  presence  of   spectroscopic  subsystems.   Finally,  the  common
distance  of both components  (flag 'p'), estimated  usually from
photometry, is an additional indicator of physical relationship.  When
accurate parallaxes of both components are measured, this criterion is
very reliable.   Real physical pairs usually  fulfill several criteria
simultaneously.

Very  wide pairs with  common PM  are not  necessarily gravitationally
bound.  Instead,  they can be  members of moving groups  or dissolving
clusters.  There  is no clear distinction between  bound and co-moving
pairs, at least observationally. In the MSC, we consider binaries with
periods   longer  than   $\sim$2\,Myr  (separations   $>$20~kau)  as
potentially unbound.

\subsection{Grading}
\label{sec:grades}

The amount and  quality of the information on  hierarchical systems in
the  MSC  is  variable;  even  the  existence  or status  of  some
companions  is uncertain.   In the  new MSC  we introduce  the grading
system  analogous to the  grades assigned  traditionally to  visual and
spectroscopic orbits.  The  grades are found in the  {\tt comp} table,
for  primary components  only.   The integer  grade  numbers have  the
following meaning.

0 -- The grade is not assigned (all secondary components have grade zero).

1 --  Not a  hierarchical system (e.g.  a simple binary  with false
claims of additional components).

2 -- Either the triple nature is doubtful, or the widest companion has
a very long period (typically longer than 2\,Myr), being for example a
co-moving  member  of  a  young  group rather  than  a  genuine  bound
companion on  a Keplerian orbit.  

3  --  Some periods  are  not known  (e.g.  a  binary discovered  from
astrometric acceleration) or the distance is uncertain.

4 -- Certainly hierarchical systems with all periods known or
estimated.  

5 -- Good-quality  systems with distance accuracy of  better than 10\%
and at least one known orbit.

There are 148  systems of grade 1 (simple  binaries), 344 questionable
hierarchies of grade  2, 252 systems of grade 3, 1168  of grade 4, and
705 of grade 5. The total number of hierarchical systems with grades 3
and above is 2125. These numbers change when the MSC is updated.

\section{The zoo of multiple stars}
\label{sec:zoo}

A  catalog  like  MSC  always  contains some  unusual  
objects.   Although  the  MSC  is  burdened  by  large  and  uncertain
discovery biases, the  presence of rare systems in  the catalog proves
their  existence in  the nature.   In this  Section we  highlight some
interesting classes of hierarchical systems. 

\subsection{Sextuples and seventuples}
\label{sec:6}

The updated MSC  contains 17 systems with six  components (four of them
are  trapezia) and four  systems with  seven  components (including  one
trapezium).   High-order   hierarchies  are  difficult   to  discover,
therefore these numbers should not  be used to evaluate their true relative
frequency.

Among   the  three   hierarchies  with   seven  components,   none  is
certain.  The most  reliable  case is  J11551+4629  (65~UMa), but  its
widest  pair  A,D at  63\farcs2,  although  definitely related,  may
represent two  subsystems in a moving group,  rather than a
genuine bound binary (its  estimated period is 0.5\,Myr). The subsystem
Da,Db was resolved  only once in 2009 and has  not been confirmed yet.
The 65~UMa  system is unique  in having four hierarchical  levels: the
wide  CPM pair  has a  3\farcs9  visual subsystem  which contains  the
641-day spectroscopic  binary with  a 1.73-day eclipsing  primary 
\citep{Zasche2012}.

Some sextuples  also have  either unconfirmed subsystems  or uncertain
status,  but several  sextuples  are genuine.   One  of those,  Castor
(J07346+3153),  is featured  in  Figure~\ref{fig:Castor}.  The  system
J04357+1010 (88~Tau) has the same hierarchical structure as Castor and
is  also   certainly  sextuple.   The   young  sextuple  J00315$-$6257
($\beta$~Tuc)  contains  only  resolved subsystems  (no  spectroscopic
binaries); it belongs to the Tucana moving group, making the status of
its widest   544\arcsec ~pair  (period $\sim$1 Myr)  questionable; it
can be just a pair of the moving group members.

The fact that many high-order hierarchies are members of moving groups
may  be significant.   Compared to  the  field, moving  groups have  a
larger            multiplicity           fraction           $\epsilon$
\citep[e.g.][]{Elliott2016,wide}.    The   frequency  of   hierarchies
independently assembled  from  $N$   subsystems  is  proportional  to
$\epsilon^N$, hence, for large $N$, they should be produced predominantly in
the high-multiplicity environment.  Of course, assembly of hierarchies
from independent subsystems is just a hypothesis.

\begin{figure}
\epsscale{1.0}
\plotone{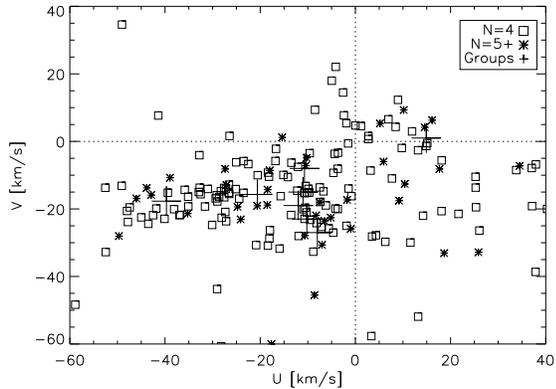}
\caption{Kinematics of high-order hierarchies. Large crosses show  Galactic
  velocities of moving groups.
\label{fig:kine}  }
\end{figure}

\begin{deluxetable}{l c ccc }[ht]
\tabletypesize{\scriptsize}
\tablewidth{0pt}
\tablecaption{Kinematics of multiple systems within 100\,pc \label{tab:UVW}}
\tablehead{
\colhead{Sample} &
\colhead{$N$} &
\colhead{$\sigma_U$} &
\colhead{$\sigma_V$} &
\colhead{$\sigma_W$} \\
  &  & \multicolumn{3}{c}{ (km~s$^{-1}$)}
}
\startdata
All     & 866  & 37.0 & 31.1 & 35.5 \\
Triple  & 665  & 39.4 & 33.8 & 39.8 \\
Quadruple & 158& 28.4 & 19.8 & 13.3 \\
$N>4$    & 42  & 25.5 & 18.8 & 15.2 \\
$P_{\rm out} > 10^7$d & 321 & 37.5 & 26.6 & 17.0 \\  
$P_{\rm out} < 10^4$d & 61 & 36.7 & 29.2 & 16.9 
\enddata
\end{deluxetable}

Using data from the {\tt  comp} table, we computed Galactic velocities
$U$, $V$, $W$ for 866 hierarchies  of grade 3 or higher located within
100\,pc from the Sun.  Figure~\ref{fig:kine} illustrates kinematics of
high-order  hierarchies  and  their  likely  association  with  moving
groups.  The  velocity dispersion  in all three  coordinates decreases
with  the increasing  multiplicity  order (Table~\ref{tab:UVW}).   One
might  think  that  high-order  hierarchies survive  predominantly  in
sparse  regions  where the  moving  groups  likely  form. However,  as
Table~\ref{tab:UVW}  shows,  the  velocity  dispersions  of  wide  and
compact  hierarchies are statistically  similar. The  average velocity
difference with the nearest moving  group for the wide ($P_{\rm out} >
10^7$ days) and  close ($P_{\rm out} < 10^4$  days) hierarchies is the
same, 24.7 and 25.9 km~s$^{-1}$, respectively.  On the other hand, the
average  velocity   difference  with  moving   groups  decreases  with
increasing  multiplicity,  from  28.6  km~s$^{-1}$ for  triples  to  17.9
km~s$^{-1}$ for hierarchies with $N>4$.

\subsection{Compact hierarchies}
\label{sec:compact}

Although  outer periods  in  triple and  higher-order hierarchies  are
typically long, there are notable exceptions. Many tertiary components
to {\it  Kepler} eclipsing  binaries have periods  of less  than three
years  \citep{Borkovits2016}.   The    Kepler  record so  far
belongs  to J19499+4137 (KIC~5897826)  with the  outer period  of only
33.9  days and the  inner period  of 1.767  days. However,  a slightly
shorter outer period of  33.07 days in J04007+1229 ($\lambda$~Tau) has
been known  since 1982; the inner  period of this B3V  triple is 3.954
days, the period ratio is  8.37.  Another compact hierarchy in the MSC
is J16073$-$2204 (HD~144548, F7V) with periods of 33.9 and 1.63 days.

Hierarchies with outer periods shorter  than three years can be discovered
relatively easily  by RVs or  by ETV.   The fact
that such systems are much less frequent compared to those with longer
periods reflects their real rareness. This has been noted in the 67-pc
sample  \citep{FG67}, where the  shortest outer  period of  1.75 year
among  $\sim$500 hierarchies belongs  to HIP~7601  (J01379$-$8259).  The
large number  of triples with  short outer periods discovered  by {\it
  Kepler}  is a  selection effect.  Although this  discovery technique
favors short outer periods, the majority of {\it Kepler} triples still
have $P_{\rm out} \ga 1000$ days.

There exist compact hierarchies  with more than three components.  The
quintuple system  VW~LMi (J11029+3025) with an  intermediate period of
355 days \citep{Pribulla2008}  contains four tightly packed solar-mass
stars and the wide companion HIP~53969 at 340\arcsec (estimated period
3\,Myr). The {\it Gaia} parallaxes of the quadruple and the companion,
8.78  and  6.10  mas  respectively,  differ  significantly,  but  this
discrepancy could arise from the photocentric motion with one year period.

\subsection{Extreme eccentricities}
\label{sec:eccentric}

Two binaries  with the largest reliably measured  eccenticity of 0.975
are members of hierarchical systems. They are J15282$-$0921 (HD 137763
or  GJ~586) and  J18002+8000 (41  Dra, HD  166866).  Both  systems are
quadruple. The inner eccentric binaries possibly have been produced by
the Lidov-Kozai cycles in hierarchies with large mutual orbit inclination
\citep{KCTF,Naoz2016}.   At such  large eccentricity,  the  orbital periods
cannot be much shorter than  $\sim$3 years without causing tidal orbit
shrinking.  So, if still larger eccentricities are to be discovered in
the future,  they will  be found in  binaries of even  longer periods.
Potential candidates for such search  can be selected from the MSC and
monitored spectroscopically to catch  (sys) short moments of passage through
the periastron.   In GJ~586, \citet{Strassmeier2013}  detected heating
of the photosphere by tides at periastron.

\subsection{Planar systems and resonances}
\label{sec:planar}

Unlike planets  in the solar system,  the orbits of  most binary stars
have  appreciable eccentricity,  while  orbits in  triple systems  are
generally  not confined  to one  common plane,  showing only  a modest
alignment \citep{ST02,moments}.  However, some hierarchies do resemble
the  solar  system  in  this  respect.   The  ``planetary''  quadruple
HD~91962 \citep{Tok2015}  consists of the  outer visual binary  with a
period  of  205\,years,  the  intermediate  9-year  spectroscopic  and
interferometric  subsystem,  and   the  inner  0.5-year  spectroscopic
binary, in a 3-tier hierarchy. The period ratios are 23 and 18.97, all
orbits have moderate eccentricity  of $\sim$0.3. The angle between the
outer and intermediate orbits is 11\degr.

The  characteristic  features  of  HD~91962 (modest  eccentricity  and
period ratio, nearly co-planar orbits)  are found in a number of other
hierarchies,      mostly      composed      of     low-mass      stars
\citep[e.g.][]{Tok2017}.    Such   {\it   planar}  hierarchies   could
plausibly be  sculpted by dissipative  evolution of their orbits  in a
viscous disk.

Another  slightly unusual  characteristic  of HD~91962  is the  integer
ratio  of 18.97$\pm$0.06   between  the intermediate  and inner
periods. This suggests a mean  motion resonance (MMR). However, a 1:19
resonance  is  very  weak  (hence unlikely)  and,  alternatively,  the
integer period  ratio could be  a mere coincidence.  In  four low-mass
triples with accurately measured period  ratios, none of the ratios is
an integer  number, although they resemble HD~91962  in other aspects
\citep{Tok2017}. So, the detection  of MMRs in stellar systems remains
controversial.   \citet{Zhu2016} claim  that three  companions  to the
eclipsing binary V548~Cyg found  by ETV have period
ratios of 1:4:12 (periods 5.5, 23.3,  and 69.9 years) and are hence in
the MMR.   However, the interpretation of eclipse  timing is sometimes
controversial,  so their result  needs confirmation,  e.g.  by  the RV
monitoring.

Resonances    are    commonly    found   in    multi-planet    systems
\citep{Fabrycky2014}.  They occur when an outer planet migrates inward
in a disk,  starts to interact dynamically with  the inner planet, and
is temporarily locked in a MMR. This mechanism can operate in stellar
systems as well.

Intriguingly,  there are  quadruple  systems where  the  ratio of  the
periods of  two inner subsystems  is a rational number.   The only
plausible interpretation is that both subsystems are in a MMR with the
outer orbit.  For example, \citet{Cagas2012} found  a doubly eclipsing
quadruple system (J05484+3057)  with periods of 1.20937  and 0.80693 days,  in a 3:2
ratio.  A similar  situation occurs  in the  massive  quadruple system
HD~5980, where the inner eclipsing binary has a period of 19.266 days,
while its tertiary  component is itself a pair with  a period of 96.56
days,  exactly 5 times  longer \citep{Koenigsberger2014}.   Both close
pairs in HD~5980 have very eccentric orbits.

\subsection{Planets in hierarchical multiple systems}
\label{sec:planets}

Planets  can orbit  single  stars  as well  as  components of  stellar
binaries and multiples.  In the latter case, a typical architecture is
a triple system where the  most massive primary component hosts one or
several planets, while the wide secondary component is a close pair of
low-mass  stars. For  example, the  5-planet system  Kepler~444  has a
companion   at  1\farcs8   which   is  a   tight   pair  of   M-dwarfs
\citep{Dupuy2016}.  Yet  another example is  94~Cet (HIP~14954), where
the  primary F8V  star hosts  a planet,  while the  secondary  pair of
M-dwarfs is  surrounded by a  dust disk \citep{Wiegert2016}.   A young
star HD~131399 in Upper Scorpius  has a planetary-mass companion Ab at
a relatively large separation of  0\farcs8 from the main component Aa,
challenging its stability in the  3\farcs2 outer system where Ba,Bb is
a  close pair  \citep{Veras2017}.  However,  it turned  out  that this
``planet'' is actually a background star \citep{Nielsen2017}.  Another
young quadruple  system in Taurus (J04417+2302)  contains three L-type
objects with masses  between brown dwarfs and planets,  while the main
star  A   has  an  estimated   mass  of  only  0.2   ${\cal  M}_\odot$
\citep{Bowler2015}.  Discoveries of similar young multiples containing
substellar bodies are likely to follow in the coming years.

The star HD~16232  has a planetary or brown dwarf  companion with $P =
335$  days, as  well as  the 0\farcs54  visual companion  discovered by
\citet{Roberts2015}.   Together  with   HD~16246  it  belongs  to  the
quadruple system J02370+2439.  Although quadruple systems with planets
are rare, their number will likely increase in the near future.

\section{Statistics}
\label{sec:stat}

As noted above, the MSC is not based on a volume-limited sample, hence
its  content does  not  reflect the  real  statistics of  hierarchical
systems;   the   statistics  are   distorted   by  the   observational
selection. Bearing this in mind, some statistical inferences can still
be  made from the  MSC. For  example, the  author compared  triple and
quadruple  systems under  the assumption  that selection  affects both
kinds of  hierarchies in a  similar way \citep{Tok2008}. The  class of
2+2  quadruple  systems,  with    components  of  similar  mass  and
comparable inner periods, similar to the $\epsilon$~Lyrae, was singled
out.   Such quadruples often  have outer  periods shorter  than $10^5$
days.

The  selection does  not depend  on  the sense  of rotation,  allowing
inferences  about relative  orientation  of orbits  in triple  systems
\citep{ST02,moments}. There is a  marked trend of co-alignment between
the angular momentum vectors of the inner and outer orbits. This trend
is  stronger  for compact  triples  with  outer  separation less  than
$\sim$50 au, but the alignment disappears for outer systems wider than
$10^3$ au.

In this Section,  the ratio of periods (or  separations) is discussed,
based on the updated MSC.

\subsection{Period ratios}
\label{sec:plps}

\begin{figure}
\epsscale{1.0}
\plotone{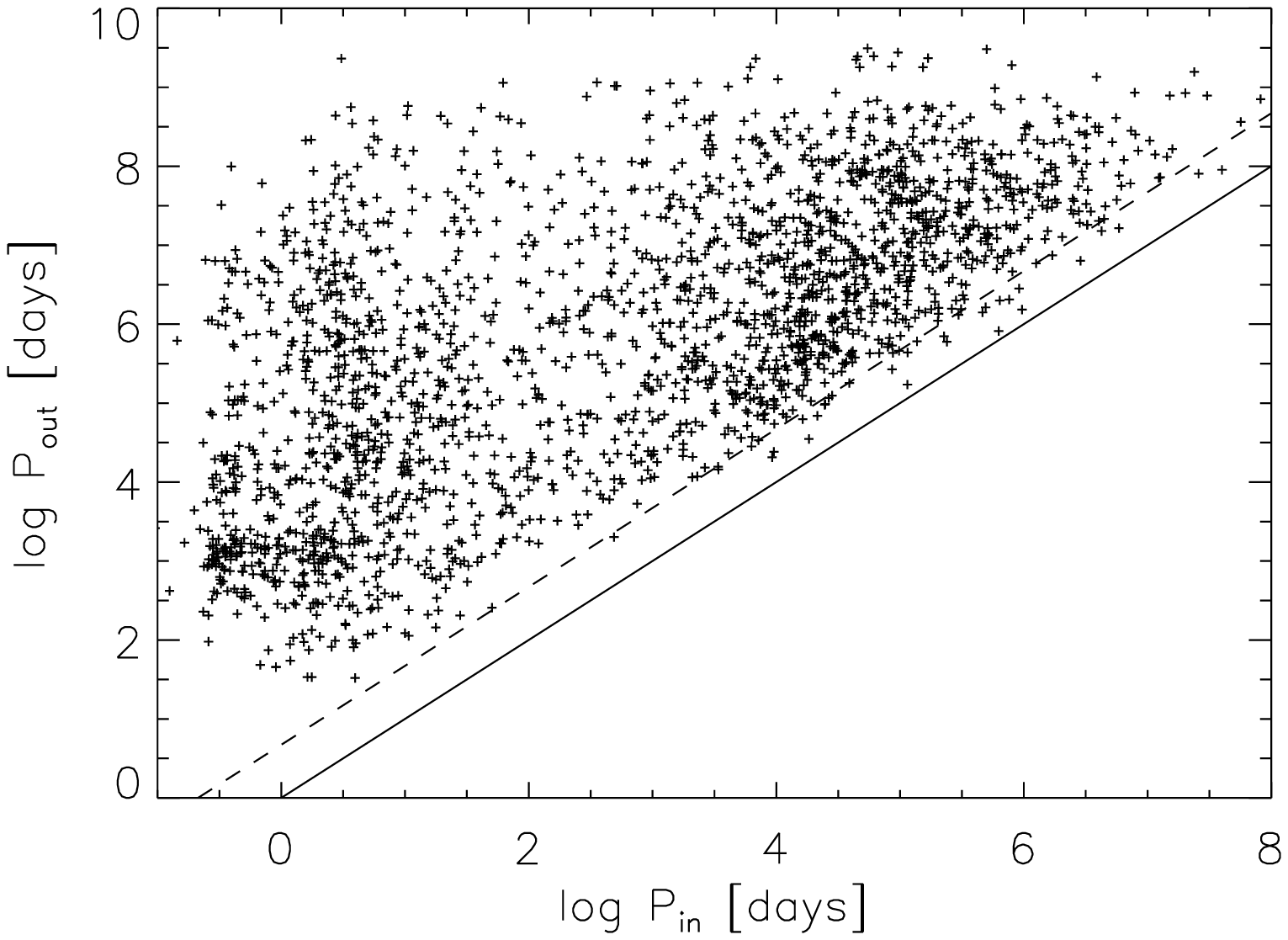}
\plotone{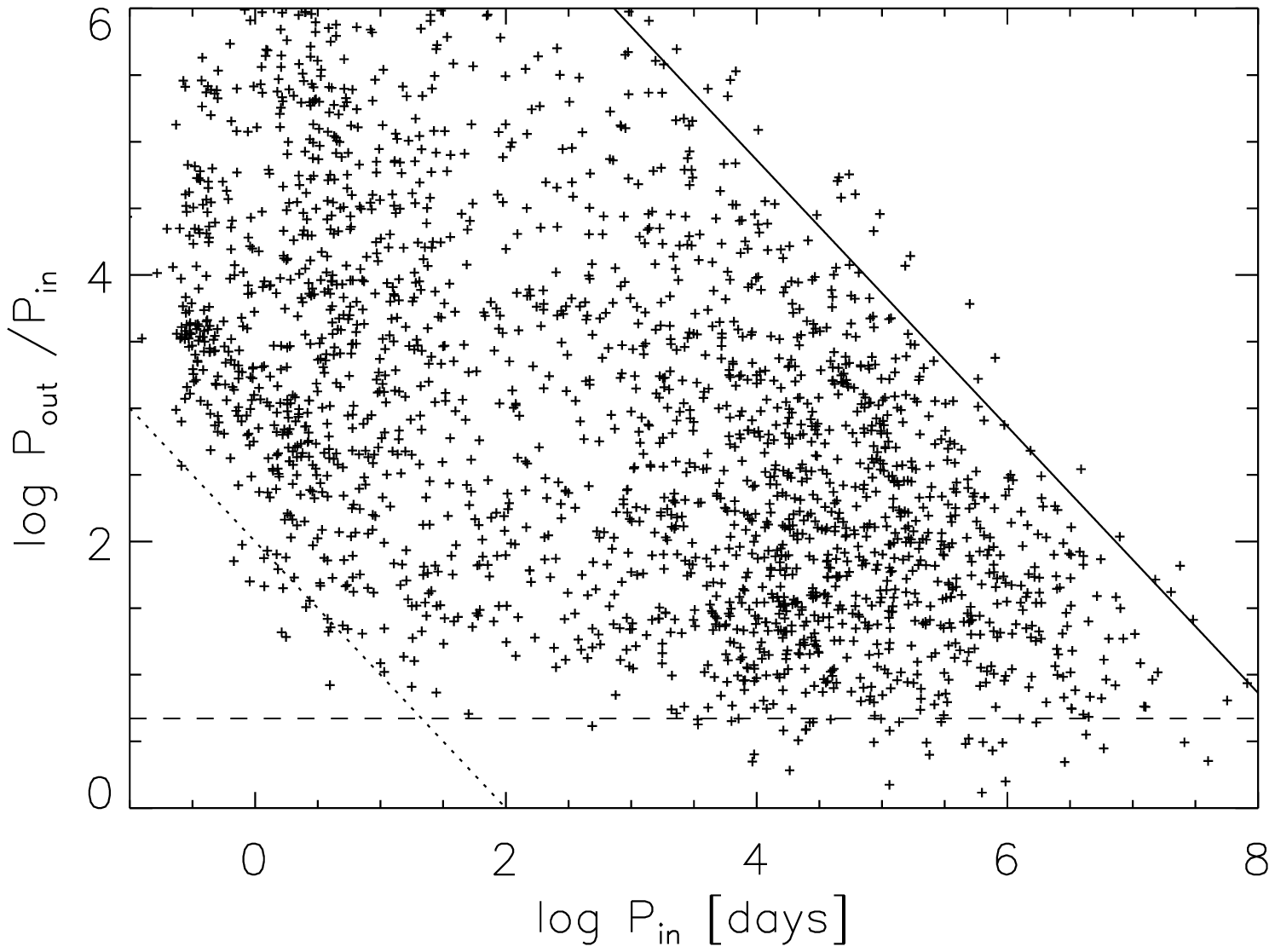}
\caption{Top: periods  at two  adjacent hierarchical levels.  The full
  line marks period  equality, the dashed line is  the stability limit
  $P_{\rm out}/P_{\rm  in} = 4.7$.  Bottom: period  ratio versus inner
  period. The dashed line is  the stability limit, the full and dotted
  diagonal  lines  mark the  outer  periods  of  2\,Myr and  100  days
  respectively.
\label{fig:plps}  }
\end{figure}

It is expected that all multiple systems (except possibly the youngest
and widest  ones) are dynamically  stable.  Several known  criteria of
dynamical  stability \citep[e.g.][]{Harrington1968}  require  that the
ratio  of  the outer  and  inner  periods,  $P_{\rm out}/P_{\rm  in}$,
exceeds  some  threshold  value   which  depends  also  on  the  outer
eccentricity  and  on  the  masses.   For example,  the  criterion  of
\citet{MA} requires  that in  stable hierarchies with  coplanar orbits
$P_{\rm out}/P_{\rm in} > 4.7$.

Figure~\ref{fig:plps}  compares periods  at two  adjacent hierarchical
levels.   It contains 2281  points (hierarchies  with more  than three
components contribute more than one point). Long periods are estimated
from the projected  separations, so they are known  within a factor of
three or so.  For this  reasons some wide triples have their estimated
period ratios  $P_{\rm out}/P_{\rm in}$ below  the dynamical stability
limit (this  issue is  treated in \S~\ref{sec:conf}).   On the
other  hand, all  hierarchies with  actually measured  orbital periods
($P_{\rm out} < 10^4$\,days) do obey the stability criterion and, with
a few  exceptions, are  elevated by  at  least $\sim$0.5  dex above  the
dashed line in Figure~\ref{fig:plps}.  This trend is better visualized
in the bottom plot  of Figure~\ref{fig:plps}.  Short inner periods are
associated  with   larger   period  ratios,  meaning   that  such
hierarchies are  more stable dynamically, compared to  the wider ones.

The  rarity of  outer  periods  shorter than  $\sim$1000  days in  the
volume-limited  67-pc sample  was noted  by \citet{FG67}  and  is also
apparent  in  the  MSC  by  the  nearly empty  lower  left  corner  in
Figure~\ref{fig:plps}.   Short  outer   periods  are  therefore  truly
uncommon.   This  feature tells  something
about  the formation mechanisms  of stellar  systems: they  likely had
larger  size at birth.   In such  case, the  short inner  periods were
produced  by subsequent  migration, while  the outer  systems migrated
less, with some exceptions discussed in \S~\ref{sec:compact}.

\subsection{Apparent configurations of wide multiple systems}
\label{sec:conf}

\begin{figure}
\epsscale{1.0}
\plotone{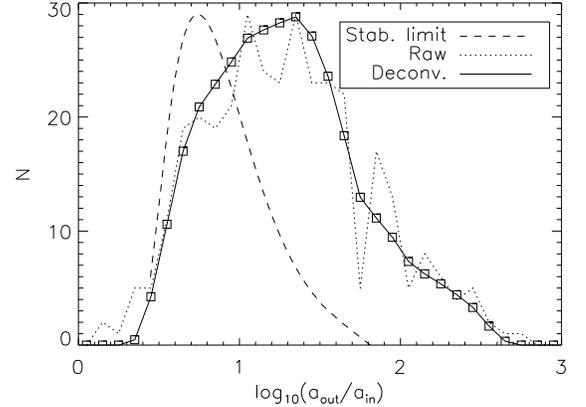}
\caption{Distribution   of   the   logarithmic  ratio   of   projected
  separations $x$ for 344 wide triple systems with $P_{\rm in} > 10^5$
  days.  The dotted  line is the histogram, the  full line and squares
  is the  distribution of  the ratio of  semimajor axes  $x_0$ derived
  from  this  histogram by  deconvolution.   The  dashed  line is  the
  assumed distribution of the critical axis ratio $\log_{10} R_0$.
\label{fig:seprat}  }
\end{figure}

\begin{figure}
\epsscale{1.0}
\plotone{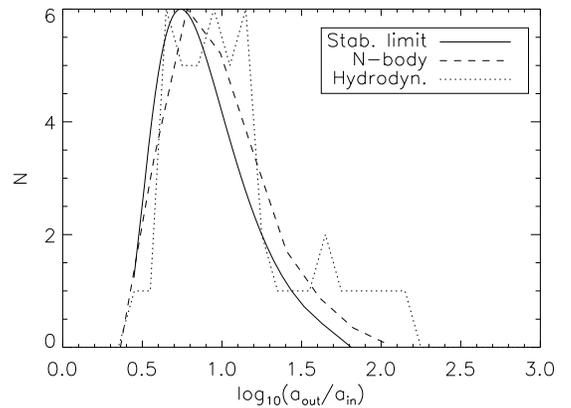}
\caption{Separation ratio distribution for triple stars produced  by
  scattering   \citep{Antognini2016}   and   in   the   hydrodynamical
  simulation of collapsing cluster \citep{Bate2014}. The solid line is
  the assumed distribution of the critical ratio $ \log R_0$.
\label{fig:theo}  }
\end{figure}

Some visual multiples  have apparently non-hierarchical configurations
with comparable separations between  their components (they are called
{\it trapezia}).  For example, the system J04519$-$3141 (HIP 22611 A,B
and  HIP~22604) contains  three F-  and G-type  stars at  separations of
99\farcs6 and  51\farcs9 that are  definitely related, based  on their
common distances, PMs, and RVs.  The orbital periods estimated from the
projected  separations, $\sim$250 and  $\sim$109 kyrs,  are comparable
and  apparently violate  the  dynamical stability  criterion (see  the
points  below  the  dashed  line in  Figure~\ref{fig:plps}).   Here  we
investigate whether such apparently  unstable triples can be explained
by  the projection effects.  As the  orbits of  wide binaries  are not
known, the issue can be studied only statistically. 

The dynamical stability criterion by \citet{MA} is $ a_{\rm
  out}/a_{\rm in}  \ge   R_0$ with
\begin{equation}
 R_0  =   2.8 (1 + q_{\rm out})^{1/15} (1 +
 e_{\rm out})^{0.4} (1 - e_{\rm out})^{-1.2} . 
\label{eq:MA}
\end{equation}
The critical ratio of semimajor axes $R_0$ depends on the eccentricity
of the outer  orbit $e_{\rm out}$ and on the  outer mass ratio $q_{\rm
  out}$. In  the following we  neglect the weak  mass-ratio dependence
and assume that  $e_{\rm out}$ has a bell-shaped  distribution $f(e) =
\pi/2 \sin(\pi e)$. This  admittedly arbitrary assumption is needed to
get an idea  of the distribution of $R_0$. Our  assumption is based on
the fact  that average eccentricity of wide  binaries containing inner
subsystems is  $ e_{\rm out} \sim  0.5 $, less than  for pure wide binaries
\citep{TK16}.  If  the outer  eccentricities  are distributed  linearly,
$f(e)=2e$, the resulting distribution of $R_0$ is broader.

We selected from  the MSC 344 wide hierarchies of grades  4 and 5 with
$P_{\rm in}  > 10^5$  days and made  a list  of their outer  and inner
separations. When the  inner orbit is known, its  semimajor axis given
in the MSC is replaced by  the actual separation from the WDS.  These
systems occupy  the upper right corner  of Figure~\ref{fig:plps}.  Let
$x  = \log_{10} (\rho_{\rm  out} /\rho_{\rm  in})$ be  the logarithmic
ratio of two  separations.  Its distribution in bins  of 0.1 dex width
is shown by the dotted line in Figure~\ref{fig:seprat}.  Although this
sample  is not  free from  observational selection,  the  discovery of
spatially  resolved wide  hierarchies with  comparable  separations is
relatively easy,  so the  leftmost part of  the histogram should  be a
reasonable approximation of the real distribution.

The  underlying distribution  of the  ratio of  semimajor axes  $x_0 =
\log_{10} a_{\rm  out} /a_{\rm in}$ is narrower  than the distribution
of $x$ which is broadened  by the projection and random orbital phases
of both  orbits.  The broadening function is  established by numerical
simulation,   assuming    a   realistic   distribution    of   orbital
eccentricities.   As  expected,   deconvolution  from  the  projection
effects  (see  Appendix)  explains  the  small  number  of  apparently
unstable triples.  The distribution of $x_0$ overlaps with the assumed
distribution of  $\log_{10} R_0$, so  many wide triples can  indeed be
close  to  the  dynamical  stability limit.   Nevertheless,  the  mean
logarithmic ratio of semimajor axes $\langle x_0 \rangle =1.33$ dex is
larger than the mean stability  limit $\langle \log_{10} R_0 \rangle =
0.93$ dex.

It is instructive to compare the separations in real wide triples with
the   theoretical   predictions.    Figure~\ref{fig:theo}  shows   the
distribution  of   $x_0$  for  hierarchical  systems   formed  in  the
hydrodynamical simulations  of a collapsing  cluster.  It uses  the data
from Table~4  of \citet{Bate2014} for  metallicities of 0.1, 1,  and 3
times solar, a total of 50 points (the 17 dynamically unstable triples
were excluded).  The  inner periods range from 100  days to $\sim$3000
years, so  these multiples are closer  than the wide  multiples in the
field considered here; still,   their distributions of $x_0$ are
quite similar.

\citet{Antognini2016}  made  a  large  series of  $N$-body  scattering
numerical experiments.  Triple  systems were produced dynamically from
binary-binary, triple-single, and triple-binary encounters.  Data from
their Figure~17 are  used to compute the distribution  of the ratio of
the  outer periastron  distance  to the  inner  semimajor axis,  $\log_{10}
[a_{\rm   out}   (1   -   e_{\rm  out})/a_{\rm   in}]$,   plotted   in
Figure~\ref{fig:theo} in  dashed line.  As this ratio  is smaller than
the  ratio of  the axes,  the  distribution of  $x_0$ for  dynamically
formed hierarchies is slightly  (by $\sim$0.3 dex) broader compared to
the dashed  line.  It matches  quite well the assumed  distribution of
the  stability  threshold  $R_0$,  illustrating the  thesis  of  those
authors that  dynamically formed hierarchies are always  ``on the edge
of  stability'' \citep[see  also][]{ST02}. Hierarchies  formed  in the
hydrodynamical  simulations  of  \citet{Bate2014}  have  $\langle  x_0
\rangle  =  1.14$  dex, being  in  this  respect  closer to  the  wide
multiples in the field, $\langle x_0 \rangle =1.33$ dex.

\section{Discussion}
\label{sec:disc}

The number of known hierarchical stellar systems has tripled since the
first MSC publication, mostly  owing to the large observational programs
conducted in the past two decades.  At the same time, the diversity of
the new MSC (e.g. the span  of primary masses) has increased and new rare
classes   of   hierarchies   were   found,  as   outlined     in
\S~\ref{sec:zoo}. Despite  these advances, the  census of hierarchical
systems remains  very incomplete; their number grows  with distance as
$d^2$, hence the apparent volume density drops as $d^{-1}$ even in the
close vicinity of the Sun. Figure~\ref{fig:dist} illustrates some
aspects of this observational selection. Thousands of hierarchical
systems within 100\,pc still wait to be discovered.

Although hierarchical  multiples can  be interesting objects  in their
own right (e.g.  to study  dynamics or to measure stellar parameters),
their  role in  revealing the  origins  of stellar  systems cannot  be
underestimated.  The  MSC helps here  by offering a  large statistical
sample  (albeit burdened  by  the selection).   The  structure of  the
period-period  diagram in Figure~\ref{fig:plps}  (bimodal distribution
of inner periods,  rarity of short outer periods,  distribution of the
period  ratios)  appears to  be  linked  to  the formation  and  early
evolution of  stellar systems.   The MSC is  also a source  of unusual
objects  highlighting  particular  aspects  of their  formation,  e.g.
hierarchies  with an architecture  resembling planetary  systems.  The
accreting    proto-triple    system    discovered   with    ALMA    by
\citet{Tobin2016}   may   eventually   become   a   nearly   coplanar
planetary-like  triple. However, further  discussion of  formation
mechanisms of binary  and multiple stars is outside  the scope of this
work.


\acknowledgements

The MSC is  based on the work of several  generations of observers who
patiently  collected data for  the benefit  of future  science, rather
than for  immediate use.  Compilation  and updates of  the binary-star
catalogs is  an often forgotten  but essential activity.   Three major
catalogs used here (WDS, VB6,  INT4) are maintained at the USNO, while
the SB9 is  hosted at the Universit\'e Libre  de Bruxelles.  This work
also  used the  SIMBAD  and  VIZIER services  operated  by Centre  des
Donn\'ees Stellaires (Strasbourg, France) and bibliographic references
from the Astrophysics Data System maintained by SAO/NASA.  Comments by
the Referee helped to improve the paper.

\appendix




\section{Deconvolution of projected separations}
\label{sec:deconv}

We simulated  a large number of  binaries with unit  semimajor axis, a
random  orbital  phase  and  orientation,  and  a  given  eccentricity
distribution.  The latter  is assumed to be either  thermal, $f(e) = 2
e$, or bell-shaped, $f(e) =  \pi/2 \sin(\pi e)$. The logarithmic ratio
of   the   separations  of  two such random  binaries,   $x   =   \log_{10}
(\rho_1/\rho_2)$,  is  distributed   symmetrically  around  zero.   An
analytical   formula  was   chosen  to   approximate   the  cumulative
distribution of $x$.  Its derivative gives the approximate form of the
broadening function due to projection, 
\begin{equation}
P(x) \propto {\mathrm e}^{-2.8 |x|} (2.8 + 2a + 2.8 a x^2)/(1 + a x^2).
\label{eq:kernel}
\end{equation}
Here  the  parameter  $a=4$  is  valid for  the  thermal  eccentricity
distribution,  while   $a=2$  gives  a  good   approximation  for  the
bell-shaped eccentricity distribution. We used $a=4$ in the following.

The histogram of  the logarithmic separation ratio $x$,  $h_i$ ($i = 1
\ldots  M$ is  the bin  number),  is related  to the  histogram of  the
logarithmic  semimajor axis  ratios $x_0$,  $H_i$, by  the convolution
equation with the kernel $P$. In discrete formulation, ${\mathbf h } =
A \cdot {\mathbf H}$, where the  $M \times M$ matrix $A$ describes the
convolution: $A_{i,j}  = C \; P(  \Delta x(i-j))$. Here  $\Delta x$ is
the bin width of the histogram and the constant $C$ normalizes the sum
of  each line to  one.  Direct  inversion of  this equation  to derive
${\mathbf H}$  from ${\mathbf  h}$ leads to  the amplification  of the
statistical noise, hence some regularization is needed.  We look for a
smooth  distribution ${\mathbf  H}$.  It  is found  by  minimizing the
merit function
\begin{equation}
{\rm min} \left[ \sum_{i=1}^M [h_i -  (A \cdot {\mathbf H})_i ]^2 + \alpha
  \sum_{i=1}^{M-1} (H_i - H_{i+1})^2 \right]
\label{eq:merit}
\end{equation}
with  a small  regularization parameter  $\alpha$. The  full  curve in
Figure~\ref{fig:seprat} was computed with $\alpha = 0.2$, using $M=30$
bins and $\Delta x = 0.1$.   The model histogram $A \cdot {\mathbf H}$
is  compatible with  the  data $\mathbf{h}$,  considering its  Poisson
statistics. Therefore the chosen parameter $\alpha$ is not too large.

\end{document}